\documentclass[final]{aipproc}
\usepackage{amsmath}
\usepackage{makeidx}
\usepackage{epsfig}
\usepackage{graphicx}

\newcommand{\be}{\begin{equation}}
\newcommand{\ee}{\end{equation}}
\def\ba{\begin{eqnarray}}
\def\ea{\end{eqnarray}}

\def\e{{\epsilon}}

\def\f{\frac}
\def\({\left(}
\def\){\right)}
\def\ls{\left[}
\def\rs{\right]}

\input{aipcheck}
\layoutstyle{8x11single} \graphicspath{{./fig3/}}

\begin{document}

\title{Canonical Lie-transform method in Hamiltonian gyrokinetics: a new
approach}
\author{Piero Nicolini\thanks{%
Piero.Nicolini@cmfd.univ.trieste.it}$\ \ ^{a,b}\ $ \ and Massimo Tessarotto%
\thanks{%
M.Tessarotto@cmfd.univ.trieste.it}$\ \ ^{a,c} $}{address={\ $^{a}
$Department of Mathematics and Informatics, University of Trieste,
Italy \\ $^{b} $National Institute of Nuclear Physics (INFN),
Trieste Section, Italy \\ $^{c} $Consortium for Magnetofluid
Dynamics\thanks{Web site: http://cmfd.univ.trieste.it}, University
of Trieste, Italy}}
\date{\today }

\begin{abstract}
The well-known gyrokinetic problem regards the perturbative expansion
related to the dynamics of a charged particle subject to fast gyration
motion due to the presence of a strong magnetic field. Although a variety of
approaches have been formulated in the past to this well known problem,
surprisingly a purely canonical approach based on Lie transform methods is
still missing. This paper aims to fill in this gap and provide at the same
time new insight in Lie-transform approaches.
\end{abstract}

\maketitle





\section{Introduction: transformation approach to gyrokinetic theory}
A great interest for the description of plasmas is still vivid in
the scientific community. Plasmas enter problems related to
several fields from astrophysics to fusion theory. A crucial and
for some aspects still open theoretical problem is the gyrokinetic
theory, which concerns the description of the dynamics for a
charged point particle immersed in a suitably intense magnetic
field. In particular, the ``gyrokinetic problem'' deals with the
construction of appropriate perturbation theories for the particle
equations of motion, subject to a variety of possible physical
conditions. Historically, after initial pioneering work
\cite{Alfen 1950,Gardner 1959,Northrop et al. 1960}, and a variety
of different perturbative schemes, a general formulation of
gyrokinetic theory valid from a modern perspective is probably due
to Littlejohn \cite{Littlejohn1979},
based on Lie transform perturbation methods \cite%
{Littlejohn1981,Littlejohn1982,Littlejohn1983,Cary et al. 1983}. For the
sake of clarity these gyrokinetic approaches can be conveniently classified
as follows (see also Fig.1):\newline
A) \textit{direct non-canonical transformation methods}: in which
non-canonical gyrokinetic variables are constructed by means of suitable
one-step \cite{Alfen 1950}, or iterative, transformation schemes, such as a
suitable averaging technique \cite{Morozov 1966}, a one-step gyrokinetic
transformation \cite{Bernstein}, a non-canonical iterative scheme \cite%
{Balescu 1986}. These methods are typically difficult (or even impossible)
to be implemented at higher orders;\newline
B) \textit{canonical transformation method based on mixed-variable
generating functions}: this method, based on canonical perturbation theory,
was first introduced by Gardner \cite{Gardner 1959,Gardner et al. 1959} and
later used by other authors \cite{Weitzner 1995}). This method requires,
preliminarily, to represent the Hamiltonian in terms of suitable
field-related canonical coordinates, i.e., coordinates depending on the the
topology of the magnetic flux lines. This feature, added to the unsystematic
character of canonical perturbation theory, makes its application to
gyrokinetic theory difficult, a feature that becomes even more critical for
higher-order perturbative calculations; \newline
C) \textit{non-canonical Lie-transform methods}: these are based on the
adoption of the non-canonical Lie-transform perturbative approach developed
by Littlejohn \cite{Littlejohn1979}. The method is based on the use
arbitrary non-canonical variables, which can be field-independent. This
feature makes the application of the method very efficient and, due to the
peculiar features for the perturbative scheme, it permits the systematic
evaluation of higher-order perturbative terms. The method has been applied
since to gyrokinetic theory by several authors \cite{Dubin1983,Lee1983,Hahm
1988,Brizard 1995};\newline
D) \textit{canonical Lie-transform methods applied to non-canonical variables%
}: see for example \cite{Hahm Lee Brizard 1988}. Up to now this
method has been adopted in gyrokinetic theory only using
preliminar non-canonical variables, i.e., representing the
Hamiltonian function in terms of suitable, non-canonical variables
(similar to those adopted by Littlejohn). This method, although
conceptually similar to the developed by Littlejohn, is more
difficult to implement.\\

All of these methods share some common features,in
particular:\newline - they may require the application of multiple
transformations, in order to construct the gyrokinetic
variables;\newline - the application of perturbation methods
requires typically the representation of the particle state in
terms of suitable, generally non-canonical, state variables. This
task may be, by itself, difficult since it may require the
adoption of a preliminary perturbative expansion.\newline An
additional important issue is the construction of gyrokinetic
canonical variables. The possibility of constructing canonical
gyrokinetic variables has relied, up to now, on essentially two
methods, i.e., either by adopting a purely canonical approach,
like the one developed by Gardner \cite{Gardner 1959,Gardner et
al. 1959}, or using the so-called ``Darboux reduction algorithm'',
based on Darboux theorem \cite{Littlejohn1979}. The latter is
obtained by a suitable combination of dynamical gauge and
coordinate transformations, permitting the representation of the
fundamental gyrokinetic canonical 1-form in terms of the canonical
variables. The application of both methods is nontrivial,
especially for higher order pertubative calculations. The second
method, in particular, results inconvenient since it may require
an additional perturbative sub-expansion for the explicit
evaluation of gyrokinetic canonical variables. \newline For these
reasons a direct approach to gyrokinetic theory, based on the use
of purely canonical variables and transformations may result a
viable alternative. Purpose of this work is to formulate a
``purely'' canonical Lie-transform theory and to explicitly
evaluate the canonical Lie-generating function providing the
canonical gyrokinetic transformation.
\begin{figure}[tbp]
\includegraphics[height=.5\textheight]{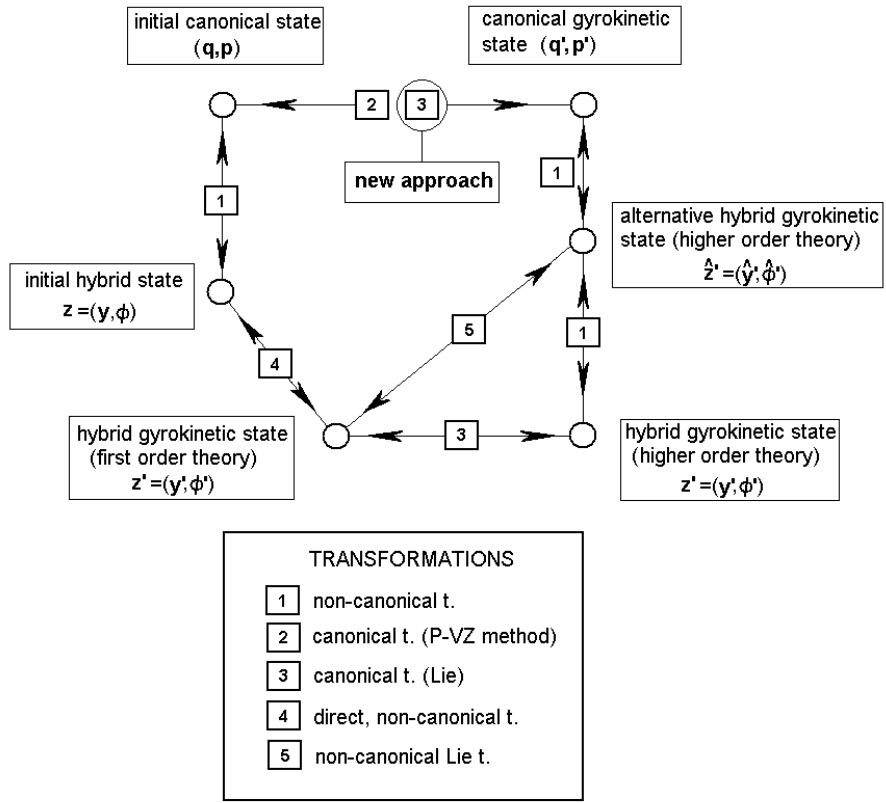}
\caption{The transformation approach to gyrokinetic theory: \fbox{2}: see
Gardner\protect\cite{Gardner 1959},\protect\cite{Gardner et al. 1959}; \fbox{%
3}: see Hahm \textit{et al.} \protect\cite{Hahm Lee Brizard 1988}
and present theory; \fbox{4}:
first obtained by Alfven \protect\cite{Alfen 1950}; \fbox{5}: see Littlejohn \protect\cite%
{Littlejohn1979}. }
\end{figure}

\section{Lie-trasform perturbation theory}
We review some basic aspects of perturbation theory for classical
dynamical systems. Let us consider the state $\mathbf{x}$ of a
dynamical system and its $d$-dimensional phase-space $M$ endowed
with a vector field $X$. With
respect to some variables $\mathbf{x}=\left\{ x^{i}\right\}$ we assume that $%
\mathbf{X}$ has representation \cite{Littlejohn1982}
\begin{equation}
\frac{dx^{i}}{dt}=X^{i}  \label{uno}
\end{equation}%
where ${\epsilon }$ is an ordering parameter. We treat all power series
formally; convergence is of secondary concern to us. By hypothesis, the
leading term $\mathbf{X}_{0}$ of (\ref{uno}) represents a solvable system,
so that the integral curves of $\mathbf{X}$ are approximated by the known
integral curves of $\mathbf{X}_{0}$. The strategy of perturbation theory is
to seek a coordinate transformation to a new set of variables $\left\{ \bar{x%
}^{i}\right\} $, such that with respect to them the new equations of motion
are simplified. Since (\ref{uno}) is solvable at the lowest order, the
coordinate transformation is the identity at lowest order, namely
\begin{equation}
\bar{x}^{i}=x^{i}+O({\epsilon })
\end{equation}
The transformation is canonical if it preserves the fundamental
Poisson brackets. It can be determined by means of generating
functions, Lie generating function or mixed-variables generating
functions, depending on the case. In the Lie transform method, one
uses transformations $T$ which are represented as exponentials of
some vector field, or rather compositions of such transformations.
To begin, let us consider a vector field $G$, which is associated
with the system of ordinary differential equations
\begin{equation}
\frac{dx^{i}}{d{\epsilon }}=G^{i}(x),  \label{due}
\end{equation}%
so that if $x$ and $\bar{x}$ are initial and final points along an
integral curve (\ref{due}), separated by an elapsed parameter
${\epsilon }$, then $\bar{x}=Tx$. In the usual exponential
representation for advance maps, we have
\begin{equation}
T=\exp ({\epsilon }G).
\end{equation}%
We will call $G$ the generator of the transformation $T$. In
Hamiltonian perturbation theory the transformation $T$ is usually
required to be a canonical transformation. Canonical
transformations have the virtue that they preserve the form of
Hamilton's equations of motion. Canonical transformation can be
represented by mixed-variable generating function, as in the
Poincare-Von Zeipel method or by means of Lie transform. In the
latter method vector fields $G$ are specified through the
Hamilton's
equations. Following a more conventional approach, we can write the (\ref%
{due}) in terms of the transformed point
\begin{equation}
\frac{d\bar{\mathbf{x}}}{d{\epsilon }}=\left[ \bar{\mathbf{x}},\omega \right]
\label{canon}
\end{equation}%
The components of the above relation are just Hamilton's equations in
Poisson bracket notation applied to the ``Hamiltonian'' (Lie generating
function) $\omega $, with the parameter ${\epsilon }$ the ``time.'' Equation
(\ref{canon}) therefore generates a canonical transformation for any ${%
\epsilon }$ to a final state $\mathbf{\bar{x}}$ whose components
satisfy the Poisson bracket condition
\begin{equation}
\left[ \bar{q}_{i},\bar{q}_{j}\right] =\left[ \bar{p}_{i},\bar{p}_{j}\right]
=0
\end{equation}%
\begin{equation}
\left[ \bar{q}_{i},\bar{p}_{j}\right] =\delta _{ij}.
\end{equation}%
To find the transformation $T$ explicitly, we introduce the
\emph{Lie operator} $ L=\left[ \omega ,\dots \right]$. Recalling
that coordinate components of vector are subject to pull back
transformation law, then one gets
\begin{equation}
\frac{dT}{d{\epsilon }}=-TL  \label{dt}
\end{equation}%
with the formal solution
\begin{equation}
T = \exp \ls -\int ^{\e} L\(\e^{\prime}\)d{\e}^{\prime} \rs .
\label{T} \end{equation}
For any canonical transformation the new
Hamiltonian $\bar{H}$ is related to the old one by
\begin{equation}
\bar{H}=T^{-1} H + T^{-1} \int^\e_0 d\e^\prime
 T\(\e^\prime\)\f{\partial \omega\(\e^\prime\)}{\partial t}. \label{ht}
\end{equation}%
To obtain the perturbation series one can expand $\omega ,L,T,H$
and $\bar{H} $ as power series in ${\epsilon }$
\begin{equation}
M=\sum_{n=0}^{\infty }{\epsilon }^{n}M_{n}  \label{emmepert}
\end{equation}%
where $M$ represents $\omega ,L,T,H$. From (\ref{dt}), equating like powers
of ${\epsilon }$, we obtain a recursion relation for the $\omega
_{n},L_{n},T_{n},H_{n}\ \ (n>0)$ which with $T_{0}=1$, gives $T_{n}$ in
terms of $L_{n}$ and $\omega _{n}$ in all orders. 

\section{The canonical Lie transform approach to gyrokinetic theory}
The customary approach based on Lie-transform methods and due to Littlejohn %
\cite{Littlejohn1979} adopts ``hybrid'' (i.e., non-canonical and
non-Lagrangian) variables to represent particle state, i.e., of the form $%
\mathbf{z=(y,}\phi \mathbf{)}$. There are several reasons, usually invoked
for this choice.  In the first place, the adoption of hybrid variables may
be viewed, by some authors, as convenient for mathematical simplicity.
However, the subsequent calculation of canonical variables (realized by
means of Darboux theorem) may be awkward and give rise to ambiguities issues %
\cite{Weitzner 1995}. Other reasons may be related to the the
ordering scheme to adopted in a canonical formulation: in fact, in
gyrokinetic theory,  the vector potential $\mathbf{A}$ in the
canonical momentum must be regarded of order $1/O(\varepsilon )$
while keeping the linear momentum of
zero order$,$ i.e., $\mathbf{p=}m\mathbf{v+}\frac{q}{\varepsilon c}\mathbf{A.%
}$ As a consequence, in a perturbative theory $\mathbf{p}$ must be expanded
retaining at the same time terms of order $1/O(\varepsilon )$ and $%
O(\varepsilon ^{0})$, a feature which may give rise to potential
ambiguities. \ According to Littlejohn \cite{Littlejohn1979} this can be
avoided by the adoption of suitable hybrid variables, which should permit to
decouple at any order the calculations of the perturbations determined by
means of suitable Lie-generators. However, a careful observation reveals
that the same ambiguity (\textit{ordering mixing}) is present also in his
method.  In fact, one finds that the first application of the non-canonical
Lie-operator method, yielding the lowest order approximation for the
variational fundamental 1-form, provides non-trivial contributions carried
by the first order Lie-generators. Probably for this reason, his approach is
usually adopted only for higher-order calculations where ordering mixing
does not appear.

In this paper we intend to point out that canonical gyrokinetic
variables can be constructed, without ambiguities, directly in
terms a a suitable canonical Lie-transform approach, by
appropriate selection of the initial and final canonical states
(see path \frame{3} in the enclosed figure),
i.e., respectively $\mathbf{x=(}\mathbf{q},\mathbf{p}{)}$ and $\mathbf{%
X^{\prime }}=(Q_{1}^{\prime },P_{1}^{\prime },{\psi _{p}}^{\prime },P_{\psi
_{p}}^{\prime },\phi ^{\prime }\mathbf{,}P_{\phi }^{\prime }\mathbf{).}$ The
latter are, by construction, gyrokinetic, i.e., the corresponding
Hamiltonian equations of motions are independent of the gyrophase angle $%
\phi .$ We want to show that the transformation $\mathbf{x}\rightarrow
\mathbf{X}^{\prime }$ can be realized, in principle with arbitrary accuracy
in $\varepsilon ,$ by means of a canonical Lie transformation of the form:%
\begin{equation}
\mathbf{x}\rightarrow \mathbf{X}^{\prime }=\mathbf{x+\varepsilon }\left[
\mathbf{X}^{\prime },\omega \right] ,
\end{equation}%
being $\omega =\omega (\mathbf{X}^{\prime },\varepsilon )$ the corresponding
Lie generator. In order to achieve this result, we shall start demanding the
following relation between the fundamental differential 1-forms, i.e., the
initial and the gyrokinetic Lagrangians, which can be shown to be of the
form:
\begin{equation}
dt\emph{L}(\mathbf{x},\frac{d}{dt}\mathbf{x,}t)=dt\mathcal{L}(\mathbf{X}%
^{\prime },\frac{d}{dt}\mathbf{X}^{\prime }\mathbf{,}%
t)+dS_{1}+dS_{2}+dS_{3}+dS_{4}.
\end{equation}%
Here $S_{1},S_{2},S_{3},S_{4}$ are suitable dynamical gauges functions,
i.e.,
\begin{equation}
S_{1}=\varepsilon \left( \mathbf{\rho }^{\prime }\mathbf{\cdot }\frac{Ze}{%
\varepsilon c}\mathbf{A}^{\prime }\right) ,
\end{equation}%
\begin{equation}
S_{2}=\frac{1}{2}\frac{Ze}{c}\mathbf{\rho }^{\prime }\cdot \nabla ^{\prime }%
\mathbf{A}^{\prime }\cdot \mathbf{\rho }^{\prime },
\end{equation}%
\begin{equation}
S_{3}=\varepsilon \mathbf{\rho }^{\prime }\cdot m\mathbf{v,}
\end{equation}%
\begin{equation}
S_{4}=m\int dQ^{\prime }v_{Q}(\psi _{p}^{\prime },Q^{\prime
},Q_{1}^{\prime },t),
\end{equation}%
where $m$ and $Ze$ are respectively the mass, the electric charge of the
particle and $\varepsilon \mathbf{\rho }^{\prime }=-\varepsilon \frac{%
\mathbf{w}^{\prime }\mathbf{\times b}^{\prime }}{\Omega ^{\prime }}$ the
Larmor radius. Moreover, $\mathbf{w}^{\prime }$ is a vector in the plane
orthogonal to the magnetic flux line, while $\mathbf{b}^{\prime }=\mathbf{B}(%
\mathbf{r}^{\prime },t)/B(\mathbf{r}^{\prime },t),$  $\Omega ^{\prime }=%
\frac{ZeB^{\prime }}{mc}$is the Larmor frequency and finally
primes denote quantities evaluated at the guiding center position
$\mathbf{r}^{\prime }$. In particular, $v_{Q}\equiv v_{Q}(\psi
_{p}^{\prime },Q^{\prime },Q_{1}^{\prime },t)$ To the leading
order in $\varepsilon $ one can prove
\begin{eqnarray}
\mathbf{r} &=&\mathbf{r}^{\prime }+\varepsilon \mathbf{\rho }^{\prime },
\label{rho} \\
\mathbf{v} &=&\mathbf{v}^{\prime }+\mathbf{w}^{\prime }\mathbf{.}
\end{eqnarray}%
The remaining notation is standard. Thus,  up to $O(\varepsilon )$ terms,
there results
\begin{equation}
\left\{
\begin{array}{c}
\mathbf{v}^{\prime }=u^{\prime }\mathbf{b}^{\prime }+\mathbf{v}_{E}^{\prime
}, \\
\mathbf{w}^{\prime }\mathbf{=}w^{\prime }\left( \mathbf{e}_{1}^{\prime }\cos
\phi ^{\prime }+\mathbf{e}_{2}^{\prime }\sin \phi ^{\prime }\right) , \\
\phi ^{\prime }=arctg\left\{ \frac{\left( \mathbf{v-v}_{E}^{\prime }\right)
\cdot \mathbf{e}_{2}^{\prime }}{\left( \mathbf{v-v}_{E}^{\prime }\right)
\cdot \mathbf{e}_{1}^{\prime }}\right\} , \\
w^{\prime }\equiv \sqrt{2B^{\prime }\mu ^{\prime }}, \\
P_{\phi }^{\prime }\ =-\frac{mc}{Ze}\mu ^{\prime },%
\end{array}%
\right.   \label{last}
\end{equation}%
where $\mathbf{v}_{E}^{\prime }=c\mathbf{E}^{\prime }\times \mathbf{b}%
^{\prime }/B^{\prime }$ is the electric drift velocity and
evaluated at the guiding center position and $\mu ^{\prime }$ is
the magnetic moment, both evaluated at the guiding center
position. Here we have adopted the representation of the magnetic
field by means of the curvilinear coordinates $(\psi
_{p}{}^{\prime }, Q^{\prime },Q^{\prime }_1)$ where $\psi
_{p}{}^{\prime }, Q^{\prime }$ are the Clebsch potentials
according to which the magnetic field reads $\mathbf{B}^{\prime
}=\nabla \psi _{p}^{\prime }\times \nabla Q^{\prime }$, whereas we
have introduced the covariant representation for the electric
drift velocity $\mathbf{v}%
_{E}^{\prime }=v_{\psi _{p}}^{\prime }\nabla \psi _{p}^{\prime
}+v_{Q}^{\prime }\nabla Q^{\prime }$. The gyrokinetic Hamiltonian
$\mathcal{K}(\mathbf{x}^{\prime },t)$, defined by means of
\begin{equation}
\mathcal{K}(\mathbf{X}^{\prime },t)=P_{Q_{1}}^{\prime }\frac{dQ_{1}^{\prime }%
}{dt}+P_{\psi _{p}}^{\prime }\frac{d\psi _{p}^{\prime }}{dt}+P_{\phi
}^{\prime }\frac{d\phi ^{\prime }}{dt}-\mathcal{L}(\mathbf{X}^{\prime },%
\frac{d}{dt}\mathbf{X}^{\prime }\mathbf{,}t)
\end{equation}%
reads
\begin{equation*}
\mathcal{K}(\mathbf{X}^{\prime },t)=-\Omega ^{\prime }P_{\phi }^{\prime }+%
\mathcal{T}+\frac{Ze}{\varepsilon }\Phi ^{^{\prime }}+
\end{equation*}%
\begin{equation}
+\left( P_{Q_{1}}^{\prime }+m\frac{\partial }{\partial Q_{1}^{\prime }}%
\left( \int dQ^{\prime }v_{Q}^{\prime }\right) \right) \frac{\partial }{%
\partial t}Q_{1}^{\prime }+\left[ P_{\psi _{p}}^{\prime }+m\frac{\partial }{%
\partial \psi _{p}^{\prime }}\left( \int dQ^{\prime }v_{Q}^{\prime }\right) %
\right] \frac{\partial }{\partial t}\psi _{p}^{\prime }+m_{s}v_{Q}^{\prime }%
\frac{\partial }{\partial t}Q^{\prime }.
\end{equation}%
Here $\mathcal{T}$ is the kinetic energy term, whereas canonical
momenta read
\begin{equation}
P_{Q_{1}}^{\prime }=\frac{\partial }{\partial \overset{\cdot }{Q_{1}^{\prime }}}\mathcal{L}(%
\mathbf{X}^{\prime },\frac{d}{dt}\mathbf{X}^{\prime }\mathbf{,}t)=\frac{%
\partial }{\partial \overset{\cdot }{Q_{1}^{\prime }}}\left\{ \emph{L}(\mathbf{x},%
\frac{d}{dt}\mathbf{x,}t)-\frac{dS_{1}}{dt}-\frac{dS_{2}}{dt}-\frac{dS_{3}}{%
dt}-\frac{dS_{4}}{dt}\right\} ,
\end{equation}%
\begin{equation}
P_{\psi _{p}}^{\prime }=\frac{\partial }{\partial \overset{\cdot }{\psi _{p}^{\prime }}}\mathcal{%
L}(\mathbf{X}^{\prime },\frac{d}{dt}\mathbf{X}^{\prime }\mathbf{,}t)=\frac{%
\partial }{\partial \overset{\cdot }{\psi _{p}^{\prime }}}\left\{ \emph{L}(\mathbf{x},%
\frac{d}{dt}\mathbf{x,}t)-\frac{dS_{1}}{dt}-\frac{dS_{2}}{dt}-\frac{dS_{3}}{%
dt}-\frac{dS_{4}}{dt}\right\} ,
\end{equation}%
\begin{equation}
P_{\phi }^{\prime }=\frac{\partial }{\partial \overset{\cdot }{\phi ^{\prime
}}}\mathcal{L}(\mathbf{X}^{\prime },\frac{d}{dt}\mathbf{X}^{\prime }\mathbf{,%
}t)=\frac{\partial }{\partial \overset{\cdot }{\phi ^{\prime }}}\left\{
\emph{L}(\mathbf{x},\frac{d}{dt}\mathbf{x,}t)-\frac{dS_{1}}{dt}-\frac{dS_{2}%
}{dt}-\frac{dS_{3}}{dt}-\frac{dS_{4}}{dt}\right\} .
\end{equation}%
Let us consider, for instance, the equation for $P_{\phi }^{\prime }.$ We
notice that  $\mathbf{\rho }^{\prime }$ coincides with $\mathbf{g}_{\mathbf{r%
}^{\prime }}^{(1)}$ the first order Lie generator of the transformation (\ref%
{rho}). Therefore, $P_{\phi }^{\prime }$ results:
\begin{equation}
P_{\phi }^{\prime }=-\frac{\mathbf{w}}{\Omega}\cdot p_{\mathbf{r}}+%
\frac{\mathbf{w}^{\prime }}{\Omega ^{\prime }}\cdot \left\{ \frac{Ze}{c}\mathbf{A}%
^{\prime }+\frac{1}{2}\frac{Ze}{c}\nabla ^{\prime
}\mathbf{A}^{\prime
}\cdot \mathbf{g}_{\mathbf{r}^{\prime }}^{(1)}+\frac{1}{2}\frac{Ze}{c}%
\mathbf{g}_{\mathbf{r}^{\prime }}^{(1)}\cdot \nabla ^{\prime }\mathbf{A}%
^{\prime }+m\mathbf{v}\right\} -m\Omega ^{\prime }\mathbf{g}_{%
\mathbf{r}^{\prime }}^{(1)}\cdot \mathbf{g}_{\mathbf{r}^{\prime }}^{(1)}=%
-\frac{1}{2}m\Omega ^{\prime }\mathbf{g}_{\mathbf{r}^{\prime
}}^{(1)}\cdot \mathbf{g}_{\mathbf{r}^{\prime }}^{(1)},
\end{equation}%
where, neglecting contributions of higher orders, the first term
on the r.h.s. has been evaluated at the effective position ${\bf
r}$. Thus, denoting $P_{\phi }\equiv \frac{\partial \dot{{\bf
r}}}{\partial\dot{\phi}}\cdot\frac{\partial}{\partial \dot{{\bf
r}}}L({\bf x}, \frac{d}{dt}{\bf x}, t) =
-\frac{\mathbf{w}}{\Omega} \cdot p_{\mathbf{r}}$, the equation can
be cast in the following form
\begin{equation}
P_{\phi }^{\prime }\cong P_{\phi }+\varepsilon \left[ P_{\phi
}^{\prime },\omega \right] ,
\end{equation}%
where $\omega $ \ is the phase function:
\begin{equation}
\omega=m\int dQ^{\prime }v_{Q}(\psi _{p}^{\prime },Q^{\prime },Q_{1}^{\prime },t)+\frac{Ze}{c}%
\mathbf{A}^{\prime }\cdot \mathbf{g}_{\mathbf{r}^{\prime }}^{(1)}+\frac{1}{2}%
\mathbf{g}_{\mathbf{r}^{\prime }}^{(1)}\cdot \frac{Ze}{c}\nabla
^{\prime
}\mathbf{A}^{\prime }\cdot \mathbf{g}_{\mathbf{r}^{\prime }}^{(1)}+\mathbf{g}%
_{\mathbf{r}^{\prime }}^{(1)}\cdot m\mathbf{v+}m\Omega^{\prime
}\int d\phi \mathbf{g}_{\mathbf{r}^{\prime }}^{(1)}\cdot \mathbf{g}_{\mathbf{%
r}^{\prime }}^{(1)}.  \label{f.generatrice di
Lie canonica}
\end{equation}%
In same fashion one determines $P_{Q_{1}}^{\prime }$ by the Lie
transform up to terms of order $O(\varepsilon )$
\begin{equation}
P_{Q_{1}}^{\prime }\cong P_{Q_{1}}+\varepsilon \left[
P_{Q_{1}}^{\prime },\omega \right] ,\ \ \ \ \ \ \ \ \ \ \ \ \ \ \
\ with\ \ \ \ \ \ \ \ P_{Q_{1}}\equiv \frac{\partial
\mathbf{r}}{\partial Q_{1}}\cdot p_{\mathbf{r}},
\end{equation}%
and similarly
\begin{equation}
P_{\psi _{p}}^{\prime }\cong P_{\psi _{p}}+\varepsilon \left[
P_{_{\psi _{p}}}^{\prime },\omega\right] ,\ \ \ \ \ \ \ \ \ \ \ \
\ \ \ \ with\ \ \ \ \ \ \
\ P_{\psi _{p}}=\frac{\partial \mathbf{r}}{\partial \psi _{p}}%
\cdot p_{\mathbf{r}}.
\end{equation}%
Therefore, it follows that $\omega$ is really \textit{the Lie
generating function of the canonical gyrokinetic transformation}
$\mathbf{x\rightarrow X}^{\prime }.$ The calculation of $\omega$
is the sought result. In terms of $\omega$ the purely canonical
gyrokinetic approach is realized. The procedure can be extended to
higher orders to develop a systematic perturbation theory.
\begin{theacknowledgments}
Work developed in the framework of the PRIN Research Program
``Programma Cofin 2002: Metodi matematici delle teorie
cinetiche''( MIUR Italian Ministry) and partially supported (for
P.N.) by the National Group of Mathematical Physics of INdAM
(Istituto Nazionale di Alta Matematica), (P.N) by the INFN
(Istituto Nazionale di Fisica Nucleare), Trieste (Italy) and
(M.T.) by the Consortium for Magnetofluid Dynamics, University of
Trieste, Italy.
\end{theacknowledgments}


\bibliographystyle{aipproc}
\bibliography{sample}

\IfFileExists{\jobname.bbl}{} {\typeout{}
\typeout{******************************************} \typeout{**
Please run "bibtex \jobname" to optain} \typeout{** the
bibliography and then re-run LaTeX} \typeout{** twice to fix the
references!} \typeout{******************************************}
\typeout{} }

\end{document}